\documentclass{webofc}
\usepackage[varg]{txfonts}   % Web of Conferences font
\usepackage{graphics}
\usepackage{amsmath}
\usepackage{slashed}
\newcommand{\Lagr}{\mathcal{L}}

\begin{document}
\title{Estimating Microscopic Nuclear Data by Compact Star Observations}

\author{
\firstname{Balázs Endre} \lastname{Szigeti}\inst{1,2}\fnsep\thanks{\email{szigeti.balazs@wigner.hu}} \and
\firstname{Gergely Gábor} \lastname{Barnaföldi}\inst{2}
%\fnsep\thanks{\email{barnafoldi.gergely@wigner.hu}}
\and
\firstname{Péter} \lastname{Pósfay}\inst{2}
%\fnsep\thanks{\email{posfay.peter@wigner.hu}}
\and
\firstname{Antal} \lastname{Jakovác}\inst{2}
%\fnsep\thanks{\email{antal.jakovac@wigner.hu}}
}
\institute{Institute of Physics, Eötvös University, Pázmány Péter Sétány 1/A, H-1117 Budapest, Hungary
\and
Wigner Research Centre for Physics, Konkoly-Thege Miklós út 29-33, H-1121 Budapest, Hungary. }

\abstract{%
We studied recent observation data of pulsar masses and radii of PSR J0740$+$6620, PSR J0348$+$0432, and PSR J1614$-$2230 from different measurements, based on the extended version of $\sigma $-$ \omega$ model. Throughout our analysis, we assumed that these pulsars are maximal-mass compact stars, thus we applied the core approximation. Based on the linear relation between the microscopic and macroscopic parameters of compact stars evaluated by our model, we estimated the average Landau mass $m_L = 752.46^{+49.1}_{-42.5}$ MeV and compressibility $K = 261.7^{+57.2}_{-28.0}$ MeV. 
}
\maketitle
\section{Introduction} \label{sec:sec1}

Compact star observables are depend on the internal characteristics of the dense nuclear matter. Modification of the microscopic nuclear model and its parameters can vary the measureable macroscopic properties of a pulsar as well. In our analysis we try to approach the problem from the inverse direction and attempt to estimate the properties of the nuclear matter from compact star observables, which is made more challenging by the {\sl masquarade problem}. It states that even large variation of the nuclear matter parameters and by exchanging of the interaction terms result in stars with very similar observable properties.

The effect of various interaction terms and the variation of the nuclear parameter values in the Lagrangian has been presented so far for the case of the extended $\sigma$-$\omega$ model~\cite{Posfay:2020}. Linear dependence of the mass and radius parameters on the Landau mass, $m_L$, compressibility, $K$, and nuclear asymmetry, $a_{sym}$ was observed and tested for maximal mass (MMS) neutron star scenarios. Ordering in the strengths of these properties variation has been also obtained in general for the maximum mass star (MMS): $ \Delta M_{max}(\delta m_L) \overset{10 \times}{>} \Delta M_{max}(\delta K)$. Landau effective mass values we already determined by former analysis \cite{Alwarez:2020,Barnafoldi:2020}. We would like to investigate the compressibilty $K$ by using recent maximal-mass pulsar observation data from PSR J0740$+$6620~\cite{J0740,XMNS,NICER}, PSR J0348$+$0432~\cite{J0348}, and PSR J1614$-$2230~\cite{J1614}. We estimate the mean $K$ and $m_L$ value including the uncertainties originating from the theory and data. The model and the analysis method was described in detail in Ref.~\cite{Alwarez:2020,Barnafoldi:2020}.

\newpage 
\section{The Model and the Equation of State}
\label{sec:eos}

One of the most commonly used model to describe the interior structure of the compact star  is the $\sigma $-$\omega$ model, which can be naturally extended by further interactions~\cite{Posfay:2020,Posfay:2019}. This model describes protons, electrons, and neutrons in $\beta$-equilibrium and approximates the nuclear force by introducing the $\sigma$, $\omega$, and $\rho$ meson with higher-order self-interaction terms for the $\sigma$ meson. In our study we would like to focus on the connection between macro- and microscopical parameters, which led us to estimate nuclear parameters with high precision from pulsar data based on this model. The Lagrangian of the corresponding to the  $\sigma$-$\omega$ model is, %has the following form, 
%
%==================equation===================
\begin{eqnarray}
%\begin{split}
\Lagr &=&
%
%nucleon term
 \overline{\Psi} \left(
i \slashed{\partial} -m_{N} + g_{\sigma} \sigma -g_{\omega} \slashed{\omega}  + g_{\rho} \slashed{\rho}^{a} \tau_{a}
 \right) \Psi
% electron terms
+ \overline{\Psi}_{e} \left(
i \slashed{\partial} - m_{e}
\right) \Psi_{e} - \lambda_{3} \sigma^{3} + \lambda_{4} \sigma^{4} \\
%
%boson term
&& +\frac{1}{2}\,\sigma \left(\partial^{2}-m_{\sigma}^2 \right) \sigma  
%
%vector meson term
- \frac{1}{4}\,\omega_{\mu \nu} \omega^{\mu\nu}+\frac{1}{2}m_{\omega}^2 \, \omega^{\mu}\omega_{\mu} 
%
%tensor meson terms
-\frac{1}{4} \rho_{\mu \nu}^{a} \, \rho^{\mu \nu \, a} + \frac{1}{2} m_{\rho}^2 \, \rho_{\mu}^{a} \, \rho^{\mu \, a} \ , \nonumber
\label{eq:wal_lag1}
\end{eqnarray}
%===============end of equation================
%
where $\Psi=(\Psi_{n},\Psi_{p})$ is the vector of proton and neutron fields, $m_{N}$, $m_{\sigma}$, $m_{\omega}$, $m_{\rho}$ are the masses of the nucleons and $\sigma$ is the scalar-, and $\omega$, $\rho$ are (iso-)vector mesons respectively. Moreover  $g_{\sigma}$, $g_{\omega}$, and $g_{\rho}$ are the Yukawa couplings corresponding to the nucleon-meson interactions. The kinetic terms corresponding to the $\omega$ and $\rho$ meson can be written in the following forms,
%
%==================equation===================
\begin{equation}
\omega_{\mu \nu}=\partial_{\mu} \omega_{\nu}-\partial_{\nu} \omega_{\mu}  \ \ \textrm{and } \ \
\rho_{\mu \nu}^{a}=\partial_{\mu} \rho_{\nu}^{a} - \partial_{\nu} \rho_{\mu}^{a} + g_{\rho} \epsilon^{abc} \rho_{\mu}^{b} \rho_{\nu}^{c}. \, 
\label{eq:wal_lag2}
\end{equation}

We are taking the mean-field approximation at zero temperature and finite chemical potential of this model. Although in the Lagrangian we can consider the fluctuations of the $\omega$ and $\rho$ vector meson fields, but because they are relative heavy their contribution can be neglected in the loop-integrals, only the $\omega_{0}=\omega$ and $\rho_{0}^{3}=\rho$ components gives non-zero value, due to symmetrical reasons but kinetic terms are disappeared. From this point the free energy corresponding to the model can be calculated as it is described in for example in Ref.~\cite{jakovac2015resummation}: 
%
%==================equation===================
\begin{eqnarray}
%\begin{split}
f_{T} \hspace{-0.2truecm}&=& \hspace{-0.2truecm}
%
%nucleon term
 f_{F}
\left(
m_{N}-g_{\sigma} \sigma,
\mu_{p} - g_{\omega} \omega + g_{\rho} \rho
\right)
+  f_{F} \left(
m_{N}-g_{\sigma} \sigma,
\mu_{n} - g_{\omega} \omega - g_{\rho} \rho
\right)
+ f_{F} \left(m_{e}, \mu_{e} \right)  \nonumber \\ 
%
%boson term
&&+ \hspace{0.1truecm} \frac{1}{2} m_{\sigma}^{2} \sigma^2  + U_{i}(\sigma)
%
%vector meson term
 - \frac{1}{2} m_{\omega}^2 \omega^2 
%
%tensor meson terms
 - \frac{1}{2} m_{\rho}^2 \rho^2 \, , 
 %\end{split}
\label{eq:wal_f}
\end{eqnarray}
%===============end of equation================
%
where $\mu_{p}$, $\mu_{n}$ and $\mu_{e}$ are the proton, neutron, and electron chemical potential respectively. The $f_{F}$ term describes the free energy contribution corresponding to one fermionic degree of freedom \cite{Posfay:2020}. For the fit of the free parameters of this model the nuclear saturation data were used~\cite{meng2016relativistic}, except for the Landau mass. Then after getting the optimal Landau mass value, we estimate the values for the compressibility as well. These two parameters are kept free and determined by comparing the mass radius diagrams corresponding to different values of the Landau mass to neutron star observations. The asymmetry energy plays negligible role on the mass and radius values of the maximal mass compact stars as it was pointed out in Ref.~\cite{Posfay:2020}. For the general relativistic description of the compact stars we assumed the usual static picture in spherically symmetric space time. %~\cite{norman1997compact,Haensel_book}. 
We calculated the mass-radius diagram by the well-known Tolman\,--\,Oppenheimer\,--\,Volkoff equations (TOV).
To focus our investigation on the effect of the nuclear matter we integrated the TOV equations, and we employed a stopping condition for the integration based on Ref.~\cite{Posfay:2020}: $p(r=R')=p_{0}$. Here, $p_{0}$ is chosen such a way that the integration stops at the core of the neutron star, so the integration does not take into account the effect of the crust and the low density EoS. To get a good approximation for $p_{0}$ we used the well known BPS nuclear equation of state which is used to describe the crust of neutron stars~\cite{BPS}. We used the highest pressure value for $p_{0}$ where the BPS EoS can still be considered valid.

The calculated $R'$ in this case corresponds to the radius of the neutron star core. The mass and radius data calculated this generally can be considered a conservative approximation of the neutron star parameters, however in the case of the maximum mass star the deviations from the normal case are insignificant~\cite{Posfay:2020}.

\section{Results} \label{Results}

Assuming maximal mass stars, we found linear dependence of both observables, the mass and radius of the maximal mass star, $M_{maxM}$ and $R_{maxM}$ on the $m_L$, where we used the saturated nuclear matter for the compressibility and asymmetry energy. The functions were fitted independently with $0.8\%$ and $17\%$ theoretical uncertainty, respectively, following Ref~\cite{Posfay:2020}: 
\begin{align}
    M_{maxM}[\textrm{M}_{\odot}] = 5.418 - 0.0043 \, m_L [\textrm{MeV}] \ , \label{eq:maxmass_M-lm} \\ %; S = 0.008 
    R_{maxM}[\textrm{km}] = 19.04 - 0.0104 \, m_L [\textrm{MeV}]   \ . \label{eq:maxmass_R-lm} %; S = 0.172
\end{align}

After we fixed the Landau mass by the MMS observation we can also obtain a one-parameter linear equation between the compressibility, $K$ and the mass and radius of the MMS compact star with $ \lesssim  2\%$ and $ \lesssim  14\%$ theoretical uncertainty, 
\begin{align}
M_{maxM}[\textrm{M}_{\odot}] = 1.940 + 0.000880\, K \, [\textrm{MeV}] \ , \label{eq:maxmass_M-K} \\
R_{maxM}[\textrm{km}] = 9.248 + 0.00718\,  K \, [\textrm{MeV}]  \ . \label{eq:maxmass_R-K}  
\end{align}

One can see that the $K$-equations~\eqref{eq:maxmass_M-K} and~\eqref{eq:maxmass_R-K} have positive gradient which means that increasing compressibility makes an MMS compact star more larger and massive. Using these derived equations and pulsar observation data we obtained the optimal Landau masses, and compressibility values for each compact object in Table 1. We also determined the average of Landau masses $m_L = 752.46^{+49.1}_{-42.5}$ MeV and mean compressibility  $K = 261.7^{+57.2}_{-28.0}$ MeV, which agrees well with the saturated nuclear matter data and with our recent works~\cite{Alwarez:2020,Barnafoldi:2020}. In contrary, calculations of $R_{maxM}$ by eqs.~\eqref{eq:maxmass_R-lm} and \eqref{eq:maxmass_R-K} present large uncertainties, especially in case of radius data from Ref.~\cite{NICER}, thus core approximation might not be valid, due to the softer EoS and its relevant contribution to the crust. This result in a $\gtrsim 1$ km larger radius.
\begin{table}[h!]
\begin{center}
\caption{The Landau mass, $m_L$ and compressiblity, $K$ values evaluated via eqs.~\eqref{eq:maxmass_M-lm}-\eqref{eq:maxmass_R-K} from measured pulsar masses and radii data denoted by ’$\ast$’, and assuming that these are maximal-mass neutron stars.}
\begin{tabular}{lllccc}
\hline %korábbi adatok
Ref. &  Pulsar & $M_{maxM}$[M$_{\odot}$] &$m_L$[MeV] & $K$[MeV] & $R_{maxM}$[km]\\
\hline
\hline
\cite{J0740}  & PSR J0740+6620 & 2.17$^{+0.11  \ \ast}_{-0.10}$ & 748.39$^{+63.3 }_{-57.2}$  &  351.8$^{+115\ \ }_{-84.5}$  & 11.25$^{+1.06}_{-1.04}$\\
\cite{J0348} & PSR J0348+0432 & 2.01$^{+0.04  \ \ast}_{-0.04}$ & 785.25$^{+20.0 }_{-20.3}$ &  206.4$^{+42.7\ \ }_{-20.5}$  & 10.87$^{+0.82}_{-0.80}$\\
\cite{J1614}  & PSR J1614-2230 & 1.97$^{+0.04  \ \ast}_{-0.04}$ & 794.47$^{+20.1 }_{-20.4}$ &  170.0$^{+15.5\ \ }_{-20.9}$ & 10.77$^{+0.82}_{-0.80}$\\
%\hline
\hline %NICER eredmény
\cite{NICER} & PSR J0740+6620 & 2.07$^{+0.07  \ \ast}_{-0.07}$ & 778.14$^{+15.3 }_{-15.5}$  &  278.2$^{+60.9\ \ }_{-60.8}$  & 11.01$^{+0.46}_{-0.47}$\\
\cite{NICER}  & PSR J0740+6620 & 2.64$^{+1.98}_{-0.65}$ & 639.42$^{+159}_{-125}$  &  277.3$^{+257\ \ }_{-178}$  & \, 12.39$^{+1.30 \ \ast}_{-0.98}$\\
\hline
%\hline %NICER + XMNS
\cite{XMNS} & PSR J0740+6620 & 2.08$^{+0.07  \ \ast}_{-0.01}$ & 769.12$^{+16.9 }_{-16.9}$  &  285.1$^{+54.8\ \ }_{-54.8}$  & 11.06$^{+0.41}_{-0.41}$\\
\hline
\end{tabular}
\end{center}
\end{table}

\begin{figure}[!ht] \label{fig:maxMass}
\begin{centering}
\includegraphics[width=0.51\textwidth]{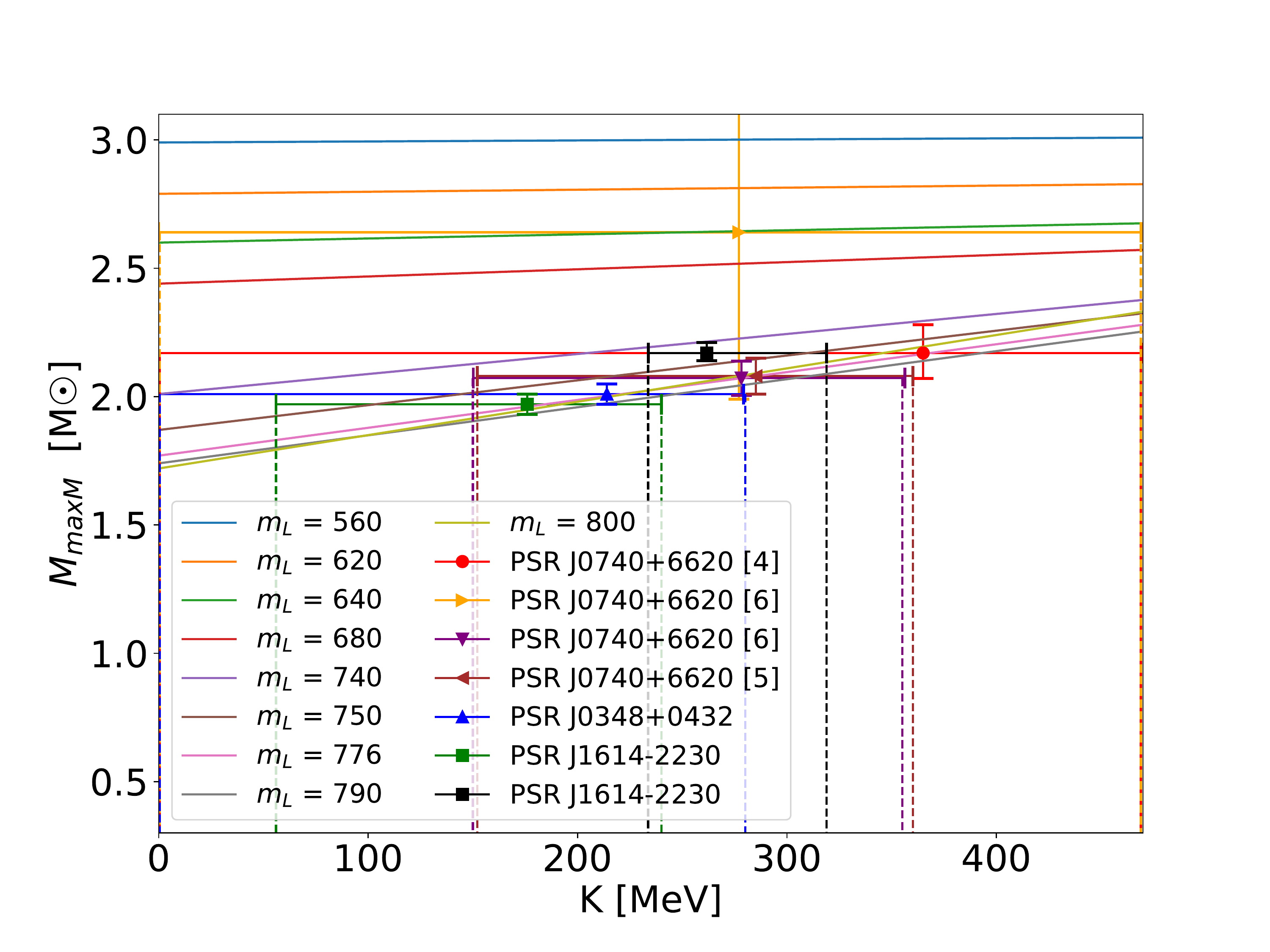} 
\includegraphics[width=0.51\textwidth]{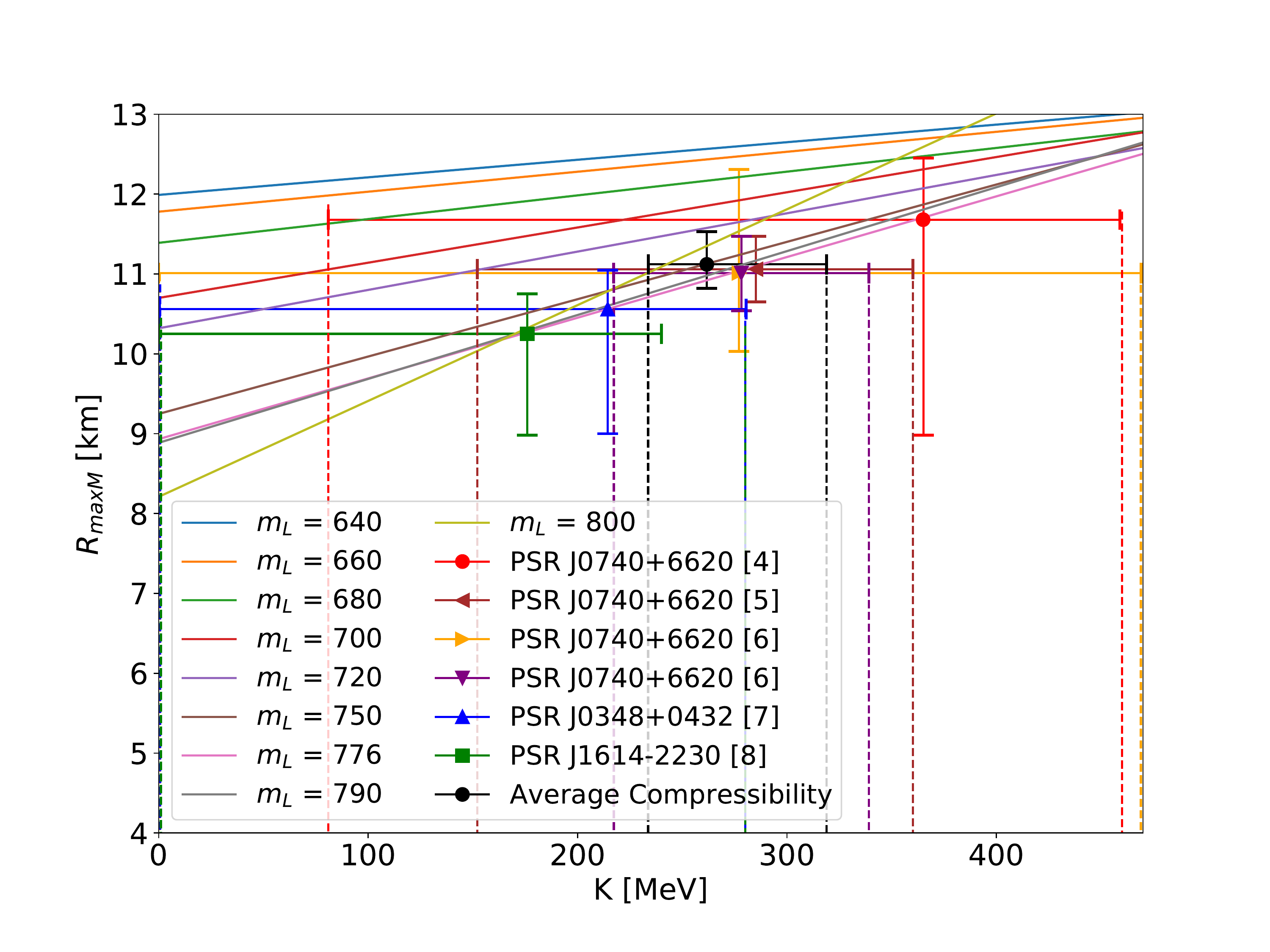} 
\end{centering}
\caption{Left panel: mass of the maximal mass star, $M_{maxM}(m_L,K)$. Right panel: radius of the maximal mass star, $R_{maxM}(m_L,K)$. Curves at both panels are as function of $K$ at various physical $m_L$ values, denoted as color lines. Comparison to pulsar data of PSR J0740$+$6620~\cite{J0740,XMNS,NICER}, PSR J0348$+$0432~\cite{J0348}, and PSR J1614$-$2230~\cite{J1614} are also plotted, including theoretical uncertainties and data errors.}
\end{figure}

On the Fig.~\ref{fig:maxMass} we plotted the projections of the $M_{maxM}(m_L,K)$ and $R_{maxM}(m_L,K)$ curves, where the $K$-dependent one-parameter linear equations are determined for various $m_L$ values. Pulsar mass and radii data from Table 1. are also plotted with color markers, where the error bars include both theoretical uncertainties from the phenomenological approach and errors from the astrophysical observation. Our results agree well with our previous results \cite{Posfay:2020,Barnafoldi:2020} and also with a Bayesian analysis carried out on this extended $\sigma$-$\omega$ model~\cite{Alwarez:2020,Barnafoldi:2020}.

\section{Summary} \label{Summary}

Macroscopical compact star observation parameters and microscopical superdenes nuclear matter properties has been formulated in Refs.~\cite{Posfay:2020,Alwarez:2020,Barnafoldi:2020}. This has been validated in the liu of the recent NICER~\cite{NICER} and NICER+XMM-Newton~\cite{XMNS} data. We find that the concept of MMS star is valid, until the core approximation stands. Microscopic nuclear parameter values: mean Landau mass $m_L = 752.46^{+49.1}_{-42.5}$ MeV and compressibility $K = 261.7^{+57.2}_{-28.0}$ MeV were given.

\section{Acknowledgement}
Authors acknowledge for NKFIH (OTKA) grants No.~K123815, K135515, PHAROS (CA16214) cost action, and for the computational resources by the Wigner GPU Laboratory.

%
%--------------------BIBLIOGRAPHY-----------------------------------------
%%%%%%%%%%%%%%%%%%%%%%%%%%%%%%%%%%%%%%%%%%%%%%%%%%%%%%%%%%%%%%%%%%%%%%%%%%%%%%%%%%%%%%%%%%%%%%%%%%%%%%%%%%%%%%
%\bibliographystyle{agsm} this is not necessary in revtex, as it sets itself
%\section*{Bibliography}

\end{document}